\documentclass[referee]{raa01}           
\usepackage{graphicx,times}
\usepackage{natbib}
\usepackage{amssymb,amsmath}
\bibpunct{(}{)}{;}{a}{}{,}

\begin{document}

   \title{Gamma/hadron segregation for ground based imaging atmospheric Cherenkov telescope using the machine learning methods: Random Forest leads}

 \volnopage{ {\bf 2012} Vol.\ {\bf X} No. {\bf XX}, 000--000}
   \setcounter{page}{1}

   \author{Mradul Sharma\inst{1} $^*$ \footnotetext{\small $*$ mradul@barc.gov.in}, J. Nayak\inst{2}, M. K. Koul\inst{1}, S. Bose
      \inst{2}  and  Abhas Mitra \inst{1}
   }

   \institute{ Astrophysical Sciences Division, Bhabha Atomic Research Centre, Mumbai, 
India; 
        \and
             The Bayesian and Interdisciplinary Research Unit, Indian Statistical Institute,
             Kolkata , India\\
\vs \no
   {\small Received ; accepted }
}

\abstract{A detailed case study of $\gamma$-hadron segregation for
{a }ground based atmospheric {C}herenkov telescope is presented. We
have evaluated and compared various supervised machine learning
methods such as {the }Random Forest{ method}, Artificial Neural
Network, Linear Discriminant method, Naive Bayes Classifiers,
Support Vector {M}achines as well as the conventional dynamic
supercut method by {simulating triggering events with} the Monte
Carlo {method and applied the results to a} {C}herenkov telescope.
It is demonstrated that the Random {F}orest method is the most
sensitive machine learning method for $\gamma$-hadron segregation.
}

   \authorrunning{M{.} Sharma et al. }            
   \titlerunning{Gamma/hadron Segregation}  
   \maketitle

%

\section{Introduction}
\label{intro}

Multidimensional datasets are very difficult to handle with conventional methods{,}
 which are generally linear in nature.
Therefore, when multidimensional data {are} encountered, the efficiency
of these methods reduces drastically as any
interdepend{e}nce among various parameters is beyond the
realm of linear methods. In the case of ground based atmospheric
Cherenkov systems, the typical characterization of {a }signal
involves more than four attributes/parameters. {P}resent day
Cherenkov systems are operating in an energy regime where
conventional methods are losing their edge on account of fading
differences among the discriminating attributes/parameter{s}
between signal and background. Therefore, the ground based gamma{ }ray
astronomy community has started exploring various options including
multivariate methods. The{se} multivariate methods fall under the
umbrella of  machine learning methods. The simplicity and intrinsic
ability of these methods to scrub out interdependence, if
any, among various attributes/parameters has made the field of
machine learning methods one of the fastest growing
scientific disciplines. These methods employ statistical tools to
decipher hidden relationship{s}, if any, among {a }few or a
collection of attributes/parameters with comparatively little
computing infrastructure.

{M}achine learning methods have been explored in the field of
ground based gamma ray astronomy for quite some{ }time. The
earliest efforts were initiated by \cite{bock}. Later on, for
$\gamma$-hadron segregation, the effectiveness of tree based
multivariate classifiers was demonstr{a}ted by two operational
ground based observatories, MAGIC (\citealt{magic}) and HESS
(\citealt{hess1,hess2,hess3}). It {should} be noted that no
machine learning method is sacrosanct as far as its superiority
over other multivariate methods{ is concerned}. Each dataset is
unique and the classifier{'s} performance is dependent on the
dataset under investigation. Therefore, in order to assess the
suitability of a classifier, each dataset needs to be  probed
independently. In this paper, we compare and evaluate various
supervised machine learning methods to assess their suitability
for $\gamma$-hadron segregation. A total of {five} machine
learning methods, namely Random Forest (RF), Artificial Neural
Network{ (ANN)}, Linear Discriminant Analysis (DISC), Naive Bayes{
(NB)} Classifier{ and} Support Vector Machine{ (SVM)} with {the
}Radial Basis Function{ (RBF)} and polynomial kernel have been
investigated. They are selected in a way to represent a type of
machine learning stream. Among these five methods, the RF method
represents {a} logic based algorithm. The {ANN} methods are
{p}erceptron based techniques. On the other hand, DISC and {NB}
Classifier are statistical learning methods. Further{more}, {SVM}
represents a rather new (1992) machine learning technique. The
signal strength after classification by each machine learning
method was compared with respect to the conventional dynamic
supercut method and a conclusion is reached to select the best
classification method.

The plan {for} the paper is as follows: In Section~2, a
brief summary of ground based atmospheric Cherenkov telescopes  and
the underlying principle will be outlined. Section~3 involves the
description of the database used to compare various machine learning
methods. The subsequent section provides an overview of all the
machine learning methods. The final two sections deal with {a}
critical analysis of all the classifiers and the conclusion
respectively.

\section{Ground based atmospheric Cherenkov systems}
\label{sect:tactic}

Ground based gamma ray astronomy is a rather new discipline. The
first successful detection of the TeV source Crab Nebula
(\citealt{crab}) took place in $1989$. With a brief lull in the
field, the next detection took place in $1992$ when the second TeV
$\gamma$-ray source Markerian $421$ (\citealt{mkn421}) was
detected and subsequently in $1996$, Mrk501 (\citealt{mkn501}) was
detected. Slowly a series of such extragalactic sources {was}
discovered. With the advent of more sensitive systems, the
catalog\footnote{ {\it http://tevcat.uchicago.edu/}} of TeV
$\gamma$-ray sources
 saw the addition of
newer sources. The present day{ field of} ground based {gamma }ray astronomy is flourishing with new detections of exotic sources.
In fact, so far more than 150 galactic and extragalactic sources have been discovered.

The detection of cosmic $\gamma$-ray sources is based on the
principle of the detection of Cherenkov photons produced by cosmic
rays in the atmosphere. When cosmic rays enter the atmosphere,
they interact with atmospheric nuclei by hadronic and
electromagnetic interaction. Electrons and the cosmic
$\gamma$-rays interact electromagnetically, i.e.{ }they generate
secondary particles by {\it {`}pair production'} and {the }{\it
{`}bremsstrahlung'} process. The hadronic cosmic rays, namely
protons and ionized nuclei{,} interact via the hadronic
interaction and also give rise to a number of secondary particles.
Such generation of secondary particles in the atmosphere is called
the {\it Extensive Air Shower}. The hadronic showers create
$\pi\degr$ particles that decay into {$\gamma$}-rays making it
difficult to distinguish {these} hadronic showers from genuine
showers{ initiated by $\gamma$-rays}. The segregation of{ showers
initiated by} $\gamma$-rays
%
is quite challenging because 
cosmic rays 
far outnumber the $\gamma$-rays by a huge margin.

\subsection{The Imaging Atmospheric Cherenkov Technique}

\begin{figure}

\vs \centering
\includegraphics[width=7.6cm]{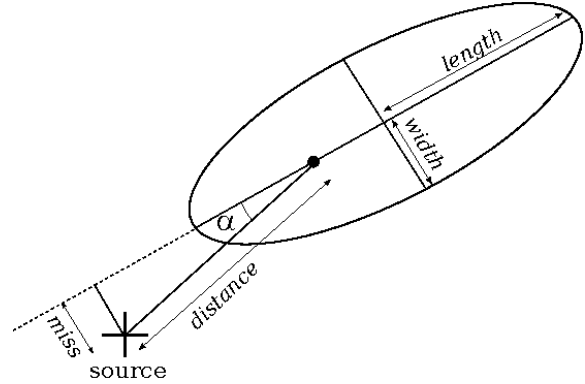}

\vs
\begin{minipage}{60mm}

\caption{{Diagram} of {a }few image parameters.
\label{figure:image}}\end{minipage}
\end{figure}

The secondary particles generated in extensive air showers move
with relativistic speeds and generate Cherenkov radiation in the
atmosphere. The technique of detecting the Cherenkov photon image
is known as the Imaging Atmospheric Cherenkov Technique (IACT). If
the shower is close enough to the telescope, the Cherenkov photons
are reflected by the telescope{'s} reflecting dish and get focused
on the camera (an array of {photomultiplier tubes} in the focal
plane of the detector). The geometrical projection of the shower
onto a detector is called an {\it image}. The IACT is used to
differentiate between $\gamma$ and hadron initiated showers on the
basis of {the }shape and orientation of the images. {The image
parameterization was introduced by {\it Hillas} and hence these
parameters are known as Hillas parameters (\citealt{Hillas}).
Image properties (analyzed offline) provide information about the
nature, energy and incoming direction of the primary particle
triggering a shower. }A representative {diagram} of {the }Hillas
parameter is shown in
Figure~\ref{figure:image}.

\section{Database used in this study}
\label{sect:database}

A database{ of Monte Carlo simulations} was generated by using the
CORSIKA air shower code (\citealt{corsika}) with the {C}herenkov
option. The simulations were carried out for the TACTIC telescope
(\citealt{mkkoul}) {located at} the Mount Abu observatory{, with
an} altitude of $\sim 1300$\,m. The{ simulated} showers were
generated at zenith angles of $5\degr$, $15\degr$, $25\degr$,
$35\degr$ and $45\degr$. The imaging camera with a total of 349
pixels was considered with the innermost $121$ pixels being used
for generating the trigger. The Cherenkov photons triggered the
telescope after encou{n}tering the wavelength dependent photon
absorption, reflection coefficient of the mirror facets, light
cone used in the camera and the quantum efficiency of
photomultiplier tubes. All the triggered events underwent the
usual image cleaning procedures described in {the }literature
(\citealt{cleaning}) to eliminate background noise.

The simulated events triggering the telescope were selected
according to the differential spectral index $2.6$ and $2.7$ for
$\gamma$ and protons respectively. The $\gamma$ events were
generated in the energy range from 1--20\,TeV. The corresponding
proton events were generated from 2--40\,TeV. In order to have a
robust and well contained image inside the camera, the
prefiltering cuts of {s}ize (photoelectrons) $\geq 50$ and a
distance cut of $0.4\degr$ $\le$  distance $\le$ 1.4$\degr$ were
applied. {This process} yielded a total of $7938$ events for{
both} $\gamma$ and protons.

\subsection{Image Parameters for Classification}
\label{subsect:image}

Various Hillas image parameters (\citealt{Hillas}) like length,
width, distance, {s}ize (photoelectron) {and }zenith angle can be
used {in the process of} $\gamma$-hadron segregation. However, the
size parameter as well as the zenith angle parameter are not
strictly separation parameters {for} {$\gamma$}-ray and hadronic
showers. In particular, the zenith angle, for instance, by itself
{cannot} be used to separate the events although different image
parameters depend on it. The same is true with the size
(photoelectron) parameter. A typical problem with these parameters
is that in case the training samples for {$\gamma$-ray} and
hadrons have a different distribution in these parameters,
 these parameters may be {considered} as separation
parameters. This may lead to a rather risky situation, 
 which
is typically handled by preparing the training samples in such a
way that their distributions on those parameters (typically {s}ize
and zenith) are as close as possible.
In this way, the uncertainty
associated with {using} such parameters as separation parameters
can be avoided.
In this study, such complexities have been
taken into account{.} In addition to these parameters, a derived
parameter {`}dens{,}' defined (\citealt{hengstebeck}) as
\begin{equation}
\rm dens=\frac{log_{10}({s}ize)} {length \times width}
\end{equation}
was also used. A total of two sets of image parameters {was}
considered. The idea was to investigate various classifiers as a
function of {the }image attributes/parameters. In the first
instance, only {five image parameters}: length,
width, distance, {s}ize and frac2 (defined as the ratio of the sum of
{the }two highest pixel signal{s} to the sum of all the
signal{s}){,} were considered from the simulation datab{a}se.
In the second case, we considered a total of {seven image parameters}. Here, in addition to the above
mentioned {five} parameters, two additional parameters{,}
namely zenith angle and dens parameter{,} were also included.
However{,} for classification purpose{s}, {the
}alpha parameter was not considered. The alpha is a very powerful
parameter as it carries the signature of the progenitor ($\gamma$ or
proton). The alpha distribution is expected to be flat for cosmic
ray protons{,} whereas it reflects a peaky behavior, for
$\leq 18\degr$ for $\gamma$-rays. In order to remove any bias of
such a strong parameter, it was not considered for classification purpose{s}. Moreover, this parameter plays a
crucial role in the estimation of signal strength. If the alpha
parameter is used in the classification{,} then the hadronic
background {cannot} be evaluated.

\section{Different classification methods}
The problem of $\gamma${-}hadron segregation is formulated as a
two class problem: $\gamma$ represents one class and the hadron is
the second class. In {the }literature, a large variet{y} of
multivariate classification methods exist{s}. However, to have a
tractable analysis, a few representative supervised machine
learning methods were selected. The classification was carried out
by using {five} different machine learning methods, namely RF,
ANN, DISC, NB Classifier and SVM with {the }RBF and polynomial
kernel. Except{ for} the RF and the Dynamic Supercut method{s},
the{ other} methods were {applied} from a commercially available
package named STATISTICA\footnote{STATISTICA {\it
http://www.statsoftindia.com/}}. On the other hand, the {RF}
method was studied by using the original {RF} code\footnote{ {\it
http://www.stat.berkeley.edu/$\sim$breiman/RandomForests/cc\_software.htm}}{.}

\subsection{Conventional Method: Dynamic Supercut}
\label{sect:dsc}

The spatial distribution of Cherenkov photons on the image plane
of the camera is parameterized on the basis of {the }shape and
{s}ize (light content) of each such image. The conventional
parameterization leads to the estimation of the image parameters
(\citealt{Hillas}). In this technique, various sequential cuts in
the image parameters are applied so as to maximize the
$\gamma$-ray like signal and reject {the }maximum number of
background events. However, this scheme has a disadvantage because
the width and length parameters  grow with the primary energy. It
is observed that {the }width and length of an image are well
correlated with the logarithm of {s}ize{; the} {s}ize of the image
provides an estimate of the primary energy. This method of scaling
the width and length parameters with the {s}ize is known as the
dynamic supercut method (\citealt{mohanty}). By employing this
method, the optimum number of cut parameters and their values
{are} estimated by numerically maximizing the so called {q}uality
factor $Q$ (\citealt{gaug}). The quality factor is defined as
\begin{equation}
Q=\frac{\epsilon_{\gamma}}{\sqrt{\epsilon_P}}\, ,
\end{equation}
where $\epsilon_{\gamma}$ and $\epsilon_P$ are $\gamma$ and hadron
acceptances respectively. The $\gamma$-acceptance is defined as the
correctly classified $\gamma$ events out of {the }total number
of $\gamma$ events and $\epsilon_P$ is the fraction of proton events
which behave like $\gamma$ events after the $\gamma$-hadron
classification. 
The image parameters {in Table~1 }lead to {the }maximum
quality factor.

\begin{table}[h]
\centering

\begin{minipage}{56mm}

\caption{Dynamic Supercut Parameters \label{tab:cuts}
}\end{minipage}

\fns\tabcolsep 5mm
\begin{tabular}{l c}
\hline\noalign{\smallskip} {Parameter}  & {Cut Value}
\\ \noalign{\smallskip}\hline\noalign{\smallskip}
Length (L)  &    0.110$\degr$ $\le$ L $\le$ $\rm (0.235 + 0.0265 \ln ({s}ize))\dg$ \\ 
Width  &   0.065$\degr$ $\le$ W $\le$ $\rm (0.085 + 0.0120 \ln ({s}ize))\dg$ \\ 
Distance (D)  & $0.4\degr$ $\le$  D $\le$ 1.4$\degr$ \\ 
Size (S)  &  S $\ge$  $50$ pe \\ 
Alpha ($\alpha$)  &  $\le$  $18\degr$  \\ 
Frac2   &  frac2 $\ge$  $0.35$ \\
\noalign{\smallskip}\hline
\end{tabular}
\end{table}

\subsection{Random Forest Method}

The RF {method} is a flexible multivariate selection method. The
algorithm for {RF} was developed by Leo Breiman and Adele
Cutler\footnote{\it
http://www.stat.berkeley.edu/$\sim$breiman/RandomForests/}.

The {RF}s are a combination of tree predictors such that each tree
depends on the values of a random vector sampled independently and
with the same distribution for all trees in the {f}orest
(\citealt{Breiman01}). The classification trees, also known as
``{d}ecision trees,{"} are machine learning prediction models
constructed by recursively partitioning the data set. Each binary
recursive partitioning splits the data sets into different
branches. The tree construction starts from the root node (the
entire dataset) and ends at {a} leaf. Every leaf node is assigned
{to} a class.
The {RF} method combines the concept of `bagging'
(\citealt{Breiman96}) and `Random Split Selection'
(\citealt{Dietterich00}).

\subsubsection{Bagging}

The RF builds on {the }bagging (\citealt{Breiman96}) technique{,}
where bagging stands for ``Bootstrapping'' and ``Aggregating''
techniques. The basic idea of bagging is to use bootstrap
re-sampling to generate multiple versions of a predictor and
combining them to make the classification. On the other hand, the
bootstrapping is based on random sampling with replacement. It
ensures that the probability of selecting an event in the sampling
(with replacement) procedure is constantly {$1/n$}. Therefore, the
probability of not selecting an event is equal to (1--$1/n$). If
the selection process is repeated $n$ times, where $n$ is very
large, the probability of not selecting an event will be $\sim
1/3$. Therefore, only 2/3 ($\sim 70 \%$) {of }events are taken for
each bootstrap sample.

\subsubsection{Random split selection}

In addition to bagging, RF also employs ``Random Split Selection''
(\citealt{Dietterich00}). At each node of the decision tree, {$m$}
variables are selected at random out of the {$M$} input vectors
and the best split is selected out of these $m$. Typically about
square root ($M$) = $m$ number of predictors are selected. Two
sources of randomness, namely random inputs and random features,
make RFs accurate classifiers. In order to measure the
classification power (separation ability) of a parameter and to
optimize the cut value, the Gini index is used{,} which measures
the inequality of two distributions. It is defined as the ratio
between (a) the area spanned by the observed cumulative
distribution and the hypothetical cumulative distribution for a
non-discriminating variable (uniform distribution, $45\degr$
line), and (b) the area under this uniform distribution.  It is a
variable between zero and one; a low Gini coefficient indicates
more equal distributions,{ while} a high Gini coefficient shows
{an }unequal distribution. \cite{Breiman01} estimated the error
rate on out-of-bag data (i.e. oob data). Each tree is constructed
on a different bootstrap sample. Since in each bootstrap training
set about one third of the instances are left out (i.e.
out-of-bag), we can estimate the test set classification error by
{applying} each case{ that is} left out of the construction of the
$t^{\rm th}$ tree {to} the $t^{\rm th}$ tree. To be precise, the
oob error estimate is the{ proportion of} misclassification {for
the} oob data.

In this study, the original {RF} code in Fortran\footnote{\it
http://www.stat.berkeley.edu/$\sim$breiman/RandomForests} was
employed and a total of $100$ trees 
was generated. The variable defined in the above code as {$m_{\rm
try}=2/3$} was taken. {Very} similar results were obtained in each
case. The resultant output of this code was compared with the
implementation of {RF} in the statistical package R\footnote{\it
http://cran.r-project.org/}. It is worth mentioning here that the
Fortran code encounter{s} some memory issues when the number of
training/test events crosses a certain threshold. However, this
limitation was not encountered in the {RF} implementat{i}on in R.

\subsection{Artific{i}al Neural Network}
\label{subsect:ann}

The ANN consists of many inputs (\citealt{ann}) which are
multiplied by weights (strength of the respective signals), and
then computed by a mathematical function {that} determines the
activation of the neuron. Another function computes the output of
the artificial neuron. The specific output demanded by the user
can be obtained by adjusting the weights of an artificial neuron.
{A }{m}ultilayer perceptron (MLP) is perhaps the most popular
network architecture in use today, due originally to Rumelhart and
McClelland (\citealt{ann1}) and discussed at length in most neural
network textbooks (\citealt{ann2}). Each neuron performs a
weighted sum of its inputs and passes it through the transfer
function to produce the output.

In this work{,} we use a{n} {MLP} network with {five} inputs, {a
}minimum {of three }hidden units {and a }maximum {of 11 }hidden
units{.} For classification tasks, the probabilistic output was
generated and {the }misclassification rate was estimated.

\subsection{Linear Discriminant Analysis}

Linear Discriminant Analysis is also known as Discriminant Function
Analysis (DFA). DFA combines aspects of multivariate analysis of
variance with the ability to classify observations into known
categories. It is a multivariate technique which {is }not only
{utilized} in classification but also estimates how good the
classification is. In this method, the discrimination functions like
canonical correlations are constructed and each function is
assessed for significance. The estimation of the significance of a
set of discriminant functions is computationally identical to
multivariate analysis of variances. After estimating the
significance, one proceeds for classification. It generally turns
out that first one or two functions play an important role while the
rest can be neglected. Each discrimination function is orthogonal to
the previous function.

In the present case, it is known that each class belongs to either
$\gamma$ or hadron; thus, {the }{\it a priori}
probabilit{ies} of these classes are known. Accordingly,
in this work, the prior probabilities are taken for classification.

\subsection{Naive Bayes Classifier}

Bayesian classifiers gained promin{e}nce in {the }early nineties
and perform very well (\citealt{langley, friedman}).  A Naive
Bayes classifier is a generative classifier technique based on the
concept of probability theory. The Bayes theorem plays a critical
role in probabilistic learning and classification. The Bayes
theorem states that
\begin{equation}
p(B/A) = \frac{p(A/B)p(B)}{p(A)}\, ,
\end{equation}
where $p(A)$ = Independent probability of $A$, $p(B)$ =
Independent probability of $B$, $p(A/B)$ = Conditional probability
of $A$ given $B$, $p(B/A)$ = Conditional probability of $B$ given
$A$, i.e. the posterior probability.

In ``Naive Bayes Classification{,}'' the different
variables/attributes/features are assumed to be strongly (naive{ly})
independent, i.e.,
\begin{equation}
p(
\langle x_1, x_2, x_3. . . x_n
\rangle |y) = \displaystyle\prod_{i=1}^n \Pi ({x_i |y})\, .
\end{equation}
Using the strong ``independence assumption'' and the prior
probabilities, the most probable class for a given $x$ is estimated.
The best class is the most likely or {\it {maximum a posteriori}} (MAP) class. The MAP estimate
gives
\begin{equation}
\displaystyle\arg \max_{\substack{B}} \ p(B/A) = \displaystyle\arg
\max_{\substack{B}} p(A/B) ~ p(B)\, .
\end{equation}
The training and evaluation from this method is very fast but the
assumption of strong independence among parameters is a
condition generally not satisfied in real world problem{s}.

\subsection{Support Vector Machine}

The SVM was introduced by \cite{boser}. It is based on the concept
of decision planes termed hyperplanes. The{se} {h}yperplanes are
constructed in multidimensional space for classification. The
decision planes separate the classes. The linear decision plane is
too limited in its application because of {the }heterogeneous
nature of experimental data. In such a case{,} the linear decision
plane lacks the ability {to perform} classification. Here
nonlinear classifiers based on {the }kernel {function} play 
 a vital role.
The kernel function (mathematical function) maps the data into
{a }higher dimensional hyperplane (feature space), where each
coordinate corresponds to one feature of the data items. In this
way, the data {are} transformed into a set of points in a Euclidean
space, leading to the classification.

In the present work, the RBF and {p}olynomial kernels are used. A
polynomial of degree $3$ with type $2$ classification was
employed. The parameters {$\gamma$} = $0.2$ and $\nu = 0.5$ were
considered. For the {RBF}, these parameters were $0.2$ and $0.35$
respectively.

\section{Comparison of Classification methods}

The above listed methods were employed to classify the events into
$\gamma$ and hadron cases. A total of $7938$ events of each type
was considered as described in {an} earlier section. Around $70\%$
of the events were used for training all the machine learning
methods and {the }rest of the data was used as a test sample. {The
s}ame tra{i}ning and test data {were} used by all the methods to
have a one to one correspondence in the results. After training,
the test sample was passed through the trained classifier and
prediction{s} of $\gamma$ and hadron class{es} were made. Our aim
is to {identify} the best classif{i}er. The accuracy of {the
}prediction rules can be evaluated by the Receiver Operator
Characteristic{ (ROC}) curves which are graphical technique{s}
(\citealt{fawcett}) to compare the classifiers and visualize their
performance. These curves are applied virtually in the field of
decision making, like in signal detection theory (\citealt{egan})
and more recently in the medical field (\citealt{swets}).

\subsection{Evaluation}
We are {considering} a binary classification problem where the two cases
are $\gamma$ and hadrons. For a binary classification problem, a
total of {four} outcomes are possible.
{T}wo outcomes are related to the correct
classification for the two classes and two for incorrect
classification. The True Positive (TP) class denotes the correct
classification of class $\gamma$ and True Negative (TN) class
represents the correct classification of class hadron. The False
Negative (FN) class reflects the class $\gamma$ incorrectly
classified
as class hadron and False Positive (FP) class is the incorrect classification of class hadron as class $\gamma$. 

The ROC plot is generated by using the above mentioned{ scenario}
for possible outcomes (TP, TN, FP, FN){.} The correctly classified
$\gamma$ are represented as {the t}rue positive rate (TPR),
estimated by defining it as{ in} (\citealt{fawcett}). T{he t}rue
positive rate is defined as
\begin{equation}
\rm TPR = \frac{TP} {TP + FN}\,.
\end{equation}
The  hadrons classified as $\gamma$ are represented by the False
Positive Rate{ (FPR)}, defined as
\begin{equation}
\rm FPR = \frac{FP} {FP + TN}\,.
\end{equation}
TPR and FPR can be defined in terms of {the }fraction of correctly
classified $\gamma$ and hadrons. From {E}quation{s} (6--7), it can
be shown that
\begin{eqnarray}
\rm TPR  & = & \epsilon_{\gamma}  \,, \\
\rm FPR & = & \epsilon_{\rm hadron}\, .
\end{eqnarray}
Hence the TPR is the accepted $\gamma$ fraction and the FPR is
defined as the accepted hadron fraction. The best classifier is the
one which provides {the }maximum TPR for {the }minimum FPR. It
{should} be noted that we are not generating the ROC curves in the
strict sense. The ROC curves lie between ($0,0$) and ($1,1$). In the
present study, in order to better understand the results, the hadron
rejection was plotted {o}n {a }logarithmic axis. Therefore, the ROC
plots in this study will differ {from} conventional ROC plots.

In order to find the best classifier, the decision boundary for
prediction was varied. Each decision boundary generated one point
in {the }$\gamma$-acceptance (TPR) and hadron acceptance{ }(FPR)
curve{s}. These rates were plotted and the  resultant plot is
referred to as a decision-plot. The decision-plot was generated
for each classifier. If the decision-plot {skews} towards the left
side, it indicates greater accuracy, i.e. a higher ratio of true
positive to false positive. In order to compare various
classifiers, the decision plot is generated after the
classification by all the methods. The top most plot in the
decision-plot turns out to be the best classifier because for the
same hadron acceptance, 
the upper plot gives the highest $\gamma$-acceptance.

\begin{figure}
\centering
\includegraphics[width=0.34\textwidth,angle=270
]{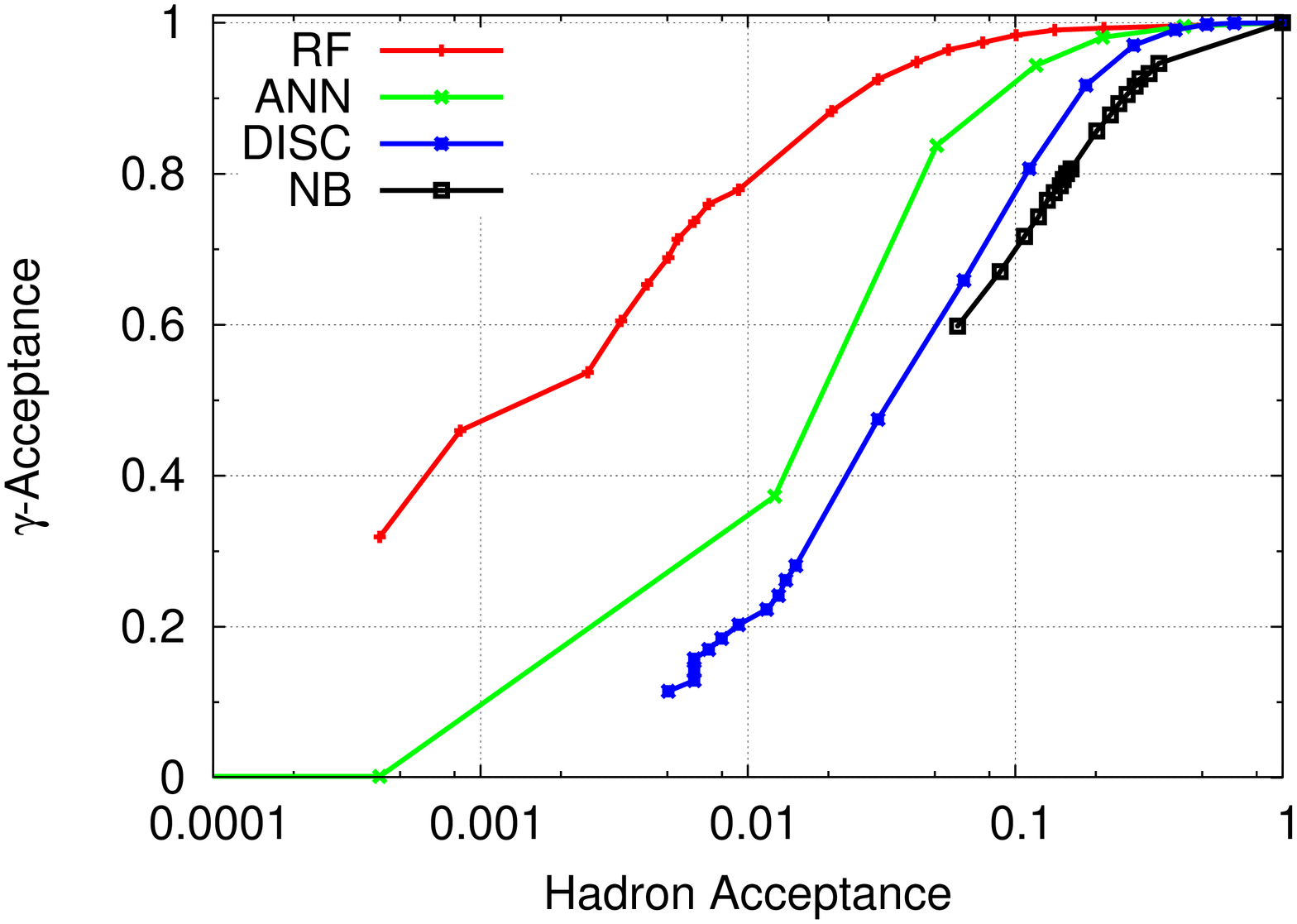}
\includegraphics[width=0.34\textwidth,angle=270
]{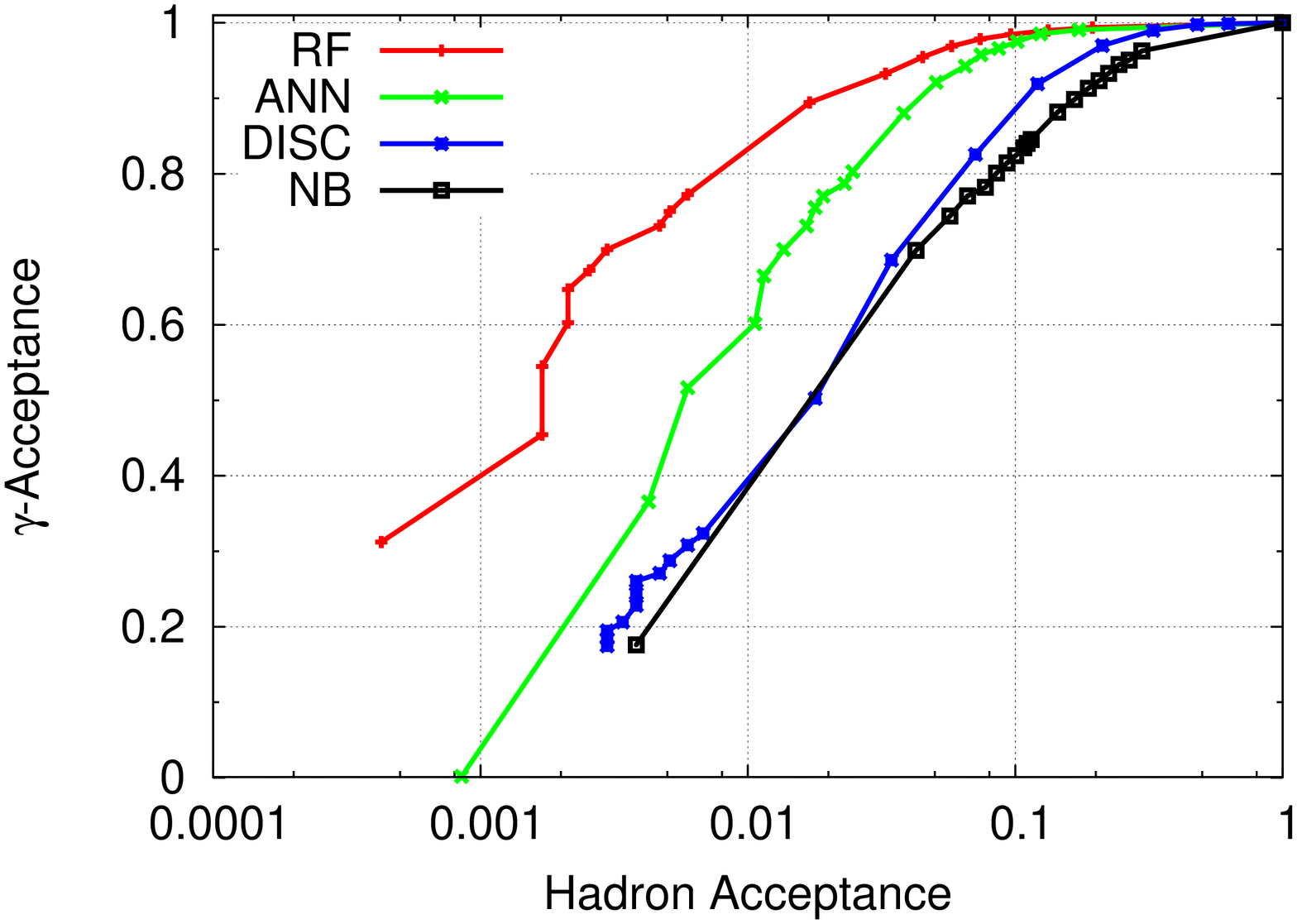}

\vspace{-3mm} \caption{\baselineskip 3.6mm \label{Figure.} Signal
vs. background acceptance. The left panel is the classification
result by using the {five} attributes/parameters. The right panel
represents it for {seven}
attributes/parameters{.}}\label{Figure:dplot}
\end{figure}

The decision{-}plot is the qualifying metric to select the most
suitable classification method. In addition to the decision-plot,
the difference among various classifiers was also quantified by
estimating the signal strength at a representative
$\gamma$-acceptance value. The  quantifying metric is designated
as ``signal strength''  {and }defined as
\begin{equation}
\sigma = \frac{S}{\sqrt{(2B + S)}}\, ,
\end{equation}
where $S=\epsilon_{\gamma}N_{\rm S}$ and $ B = \epsilon_{p} N_{\rm
B}$ (\citealt{li1983}) are the signal and background events
respectively.  The signal strength was estimated by taking
{$N_{\rm B}=10\,000$} and $N_{\rm S}=500$ (\citealt{bock}). Since
the conventional dynamic supercut method estimated the
$\gamma$-acceptance {to be} $57.4\%$, the hadron acceptance from
each classifier was derived from the decision-plot at{ a}
$\gamma$-acceptance of $57.4\%$. The decision plot was generated
for two sets of image parameters. As mentioned earlier, two sets
were considered to evaluate the classification strength as a
function of the number of image parameters. The decision-plots for
these two cases are shown in Figure~\ref{Figure:dplot}. 

\begin{figure}
\begin{center}
\includegraphics[width=0.40\textwidth,angle=270,clip]{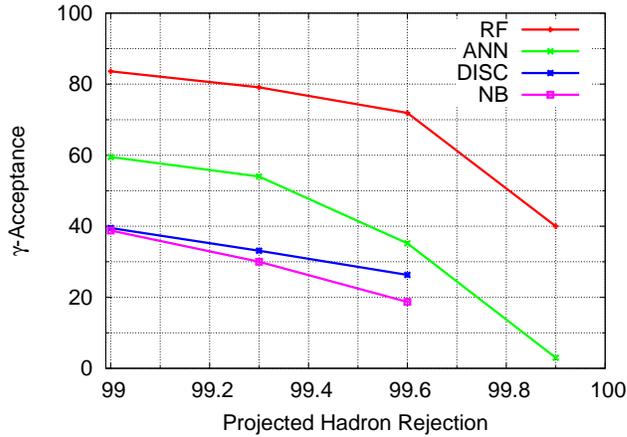}

\begin{minipage}{90mm}

\caption{\label{Figure.} {$\gamma$}-acceptance as a function of
projected hadron rejection.\label{Figure:projplot}}\end{minipage}
\end{center}
\end{figure}

The comparison of  decision-plot for {RF} methods for {five} and
{seven} set{s} of image parameters shows that{ the} {RF} method
yields {a }better classification strength. This difference in the
classification is{,} however, small and of the order of $\sim
10\%$ in the $\gamma$-ray acceptance for the given hadron
acceptance range. This difference results because of {the larger}
number of image parameters and guides us to choose more number{s}
of image parameters during the training of {the }classification
method. The decision-plot for {the }artificial neural network
method also reflects a tendency to prefer more number{s} of image
parameters for better classification. As per the decision-plot,
the {other }two methods also indicate {a} positive effect on the
classification strength {with} more number{s} of image parameters.
The decision plot provides a{n} estimate{ of} the possible
$\gamma$-acceptance for a user chosen background (hadron)
rejection. Any classifier yielding the maximum
{$\gamma$}-acceptance for a given  hadron acceptance decides the
{quality} of the classifier.
Figure~\ref{Figure:projplot} 
 shows the
$\gamma$-acceptance as a function of projected hadron-rejection for
{four} representative projected hadron-rejection
values, viz $99\% $, $99.3\%, 99.6\%${ and }$99.9\%$.

For a hadron rejection of $99.9\%$, the RF method yields $\sim 40\%$
$\gamma$-acceptance. In comparison to this, any classifier coming
closest to {RF} is ANN, which for the same
hadron rejection secures a mere $\sim 3\%$ {for} $\gamma$-acceptance.
{T}he {other }two classifiers fail to go
beyond $99.6\%$ {for} projected hadron rejection. Further{more}, they yield a
much {smaller} $\gamma$-acceptance compared to
{the }above two classifiers{,} even at a projected hadron
acceptance of $99.6\%$.

In addition to estimating the signal strength, the
misclassification rate was also estimated by using {a} confusion
matrix. The misclassification rate and the signal strength are
shown in Table~\ref{tab:mrate}. 

\begin{table}

\centering

\begin{minipage}{130mm}
\caption{\baselineskip 3.6mm Misclassification Rate and Signal
Strength with {Five}
 and
{Seven} Image Parameters\label{tab:mrate}}\end{minipage}

\fns\tabcolsep 5mm
\begin{tabular}{lcc c}
\hline\noalign{\smallskip}
                                & Misclassification Rate (\%)      & Signal Strength \\
{Classification method}  & $R_{5}/R_{7}$    &\textbf{$\sigma_{5}/\sigma_{7}$} \\
\noalign{\smallskip}\hline\noalign{\smallskip}
Random Forest                   & 5.44$/$4.43     &  15.46\,/\,15.73          \\
Automated Neural Network        & 7.40\,/\,5.82     &
9.8\,/\,13.30
\\ Dynamic Supercut method        &  ---                &  9.1\,/\,12.92
             \\ Linear Discriminant Analysis  & 12.11\,/\,10.08
&  8.10\,/\,10.37          \\ Naive Bayes Classifiers &20.57 /
14.00
&  7.8\,/\,10.32           \\
Support Vector Machine          &          &                                \\ 
(i) with RBF kern{e}l              &  9.18\,/\,16.08  &   na                          \\ 
(ii) with polynomial kernel       &  10.19\,/\,16.12 &    na
\\ \noalign{\smallskip}\hline
\end{tabular}
\end{table}

The positive effect of {a greater} number of
parameters is better {seen }by {a} quantification
of {the }misclassification rate as well as the signal strength.
Table \ref{tab:mrate}  shows that {a higher} number of
attributes/parameters for training the classifier improves the
signal strength while the misclassification rate goes down.

Such improvement in the misclassification rate as well as the
signal strength is equally visible in all the classification
methods. It {should} be noted that{ entries related to} the {SVM}
in Table~\ref{tab:mrate}  are absent. Only the misclassification
rate is given. Many classification methods (ANN, DISC, NB) used in
STATISTICA  give the probabilistic output as well as {the
}prediction probability{,} but there are instances where the
prediction is a hard prediction, i.e. in terms of YES or NO
output. In the case of SVM, the STATISTICA package yields hard
predictions, thereby hinder{ing} the generation of a set of
confusion matri{ces} for different decision boundaries. Due to the
lack of probabilistic output from SVM, it is difficult to estimate
the signal strength. However, the misclassification rate from
Table \ref{tab:mrate}  for SVM with both the kernels (RBF and
polynomial) suggest{s} that for the given dataset, {$\gamma$} and
hadron acceptance{s} will remain lower compared to {those} of {the
}RF and ANN method{s}. Based on this premise, it {can be}
concluded that the SVM will not be able to match these two
classifiers for our requirement.

Note that the strength of {the }ROC curves is generally exploited
by comparing various classifiers and {a} suitable classifier is
selected. The classifier is selected on the basis of its position
in the ROC space. The {top }left most plot is considered {to be}
the best classifier. However, this view of selecting the
class{i}fier on the basis of its position in the {top }left most
part {of} the decision-plot is over simplistic. The
Precision-Recall (PR) curves are more fundamental than the ROC
plots. According to the theorem (\citealt{davis06}), ``For a fixed
number of positive and negative examples, one curve dominates a
second curve in {the }ROC space if and only if the first dominates
the second in {the }{PR} space{.}" The precision is defined as
\begin{equation}
\rm Precision = \frac{TP}{TP + FP}\, {.}
\end{equation}
The {p}recision essentially reflects the fraction of examples
classified as positive which are truly positive, i.e. predicted
positives (here class $\gamma$). The Recall is the {TPR}. In the PR space, the recall is
plotted on the {$x$}-axis and the {p}recision is plotted on
the {$y$}-axis.

\begin{figure}

\centering

\includegraphics[width=0.34\textwidth,angle=270,clip]{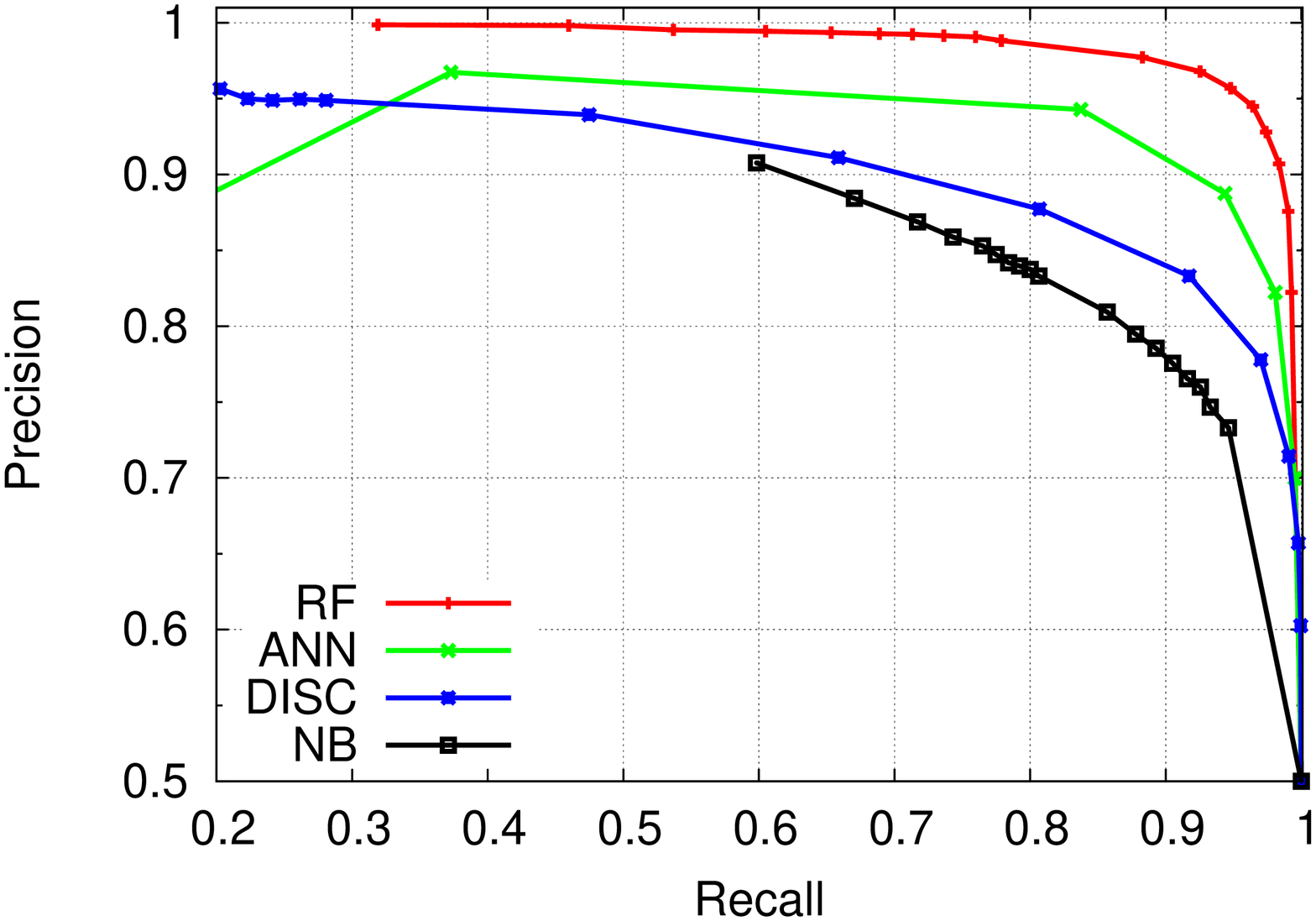}
\includegraphics[width=0.34\textwidth,angle=270,clip]{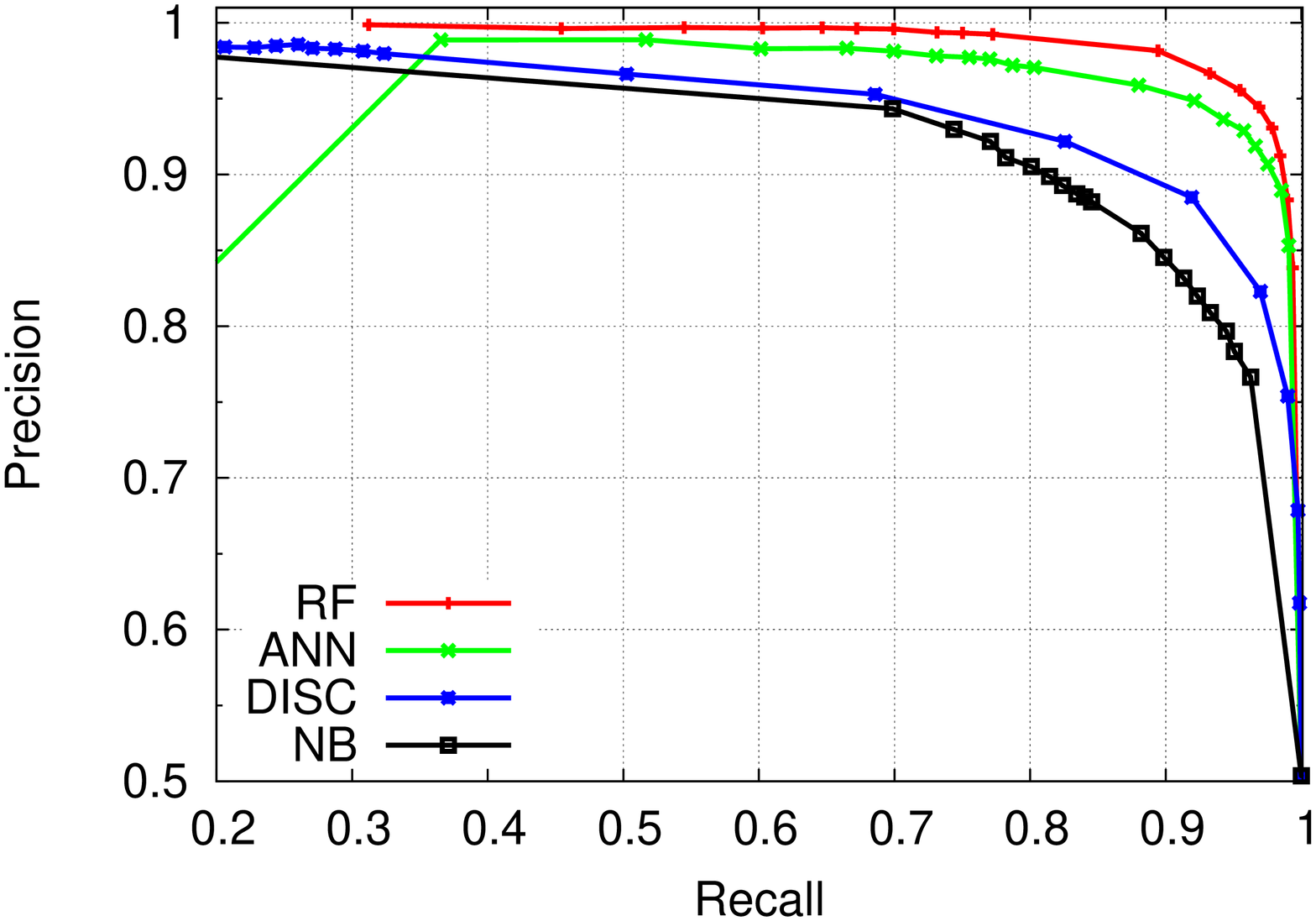}

\caption{\baselineskip 3.6mm \label{Figure.} {PR} curves. The left
panel represents the PR curve for the {five}
attributes/parameters. The right panel represents it for {seven}
attributes/parameters.}\label{fig:pr-curve}
\end{figure}

The classifier attaining the top position in the PR space and hence
in the ROC space (as per the above mentioned theorem) is {regarded}
as the best classifier. Therefore, in order to reach the conclusion
about the best classifier, it is important to evaluate the
classifier performance in the PR space. The {PR} plot is generated
for both sets of image parameters and is shown in
Figure~\ref{fig:pr-curve}. 

The {RF} method retains the top most position
in the ROC curve as well as in the PR{ }space compared to
the{ other} classifiers. Therefore, on the basis of these two
curves, it {can be} concluded that since the {RF}
method dominates all the other classifiers, for the given dataset,
it turns out to be the best classifier.  It {should be} be noted that the
superiority of {the }PR curve over ROC plots is more pronounced
when there is skew{ness} in the class distribution {of} a dataset.

\section{Conclusions}

Five different machine learning methods were evaluated and compared
to decide {which} of these methods
{is most suitable }for $\gamma$-hadron segregation. Given the
position of all the methods in the ROC space, {the }PR{ }space and
the misclass{i}fication rate for the given dataset, the  trend
reflects the superiority of {RF} and 
{ANN} compared to 
{other }methods, i.e. {the }{DISC}, {NB} Classifier and
{SVM}. The signal strength was
estimated by using {a} confusion matrix at a representative value {for}
$\gamma$-acceptance of $0.574$. This acceptance value was chosen
because the conventional dynamic supercut method yields the same
$\gamma$-acceptance. The dynamic supercut method yields a signal
strength of $\sigma_{0.574} = 12.92$, whereas the signal
strength{s} are $15.73$ and $13.30$ from the {RF} method and the {ANN} method respectively. It is clear that these two
methods yield better results compared to the conventional dynamic
supercut method. For the given dataset, {the }{RF} method gives {an }almost $20\%$ improvement in the signal
strength over the {ANN} method.
{A s}imilar story is repeated in the estimation of
{the }misclassification rate.  It is of course difficult to make
a generalized statement about the superiority of {the
}{RF} method over any other method.
Yet, the dominance of {the }{RF} method
in {the }ROC plot as well as in the
{PR} space indicates that for the given
dataset, results are tilting in favor of the {RF} method. In addition to {the }above classifying
metric, the {RF} method has an advantage in
terms of computational time over the perceptron based method{s,}
like {ANN}. As the
number of perceptrons increases, it becomes {very }computationally
expensive{; an} increase in the number of
attributes/parameters add{s} to the computational expense.
Also{,} unlike {the }{ANN} method, which acts as a black box, the {RF} method is quite easy to understand. Further{more}, the
{RF} method demands very little processing
capabilities. Finally, the {RF} method takes
care of parameters with little or no separation power{,} whereas
{ANN} performance can be severely
affected by the inclusion of such parameters.

In the next phase, a similar study will be carried out with a
bigger dataset and the best method will be employed for
$\gamma$-hadron segregation by taking experimental data. With the
ever increasing data volume and the inclusion of {larger}
number{s} of attributes/parameters in the field of ground based
$\gamma$-ray astronomy, the {RF} method, or rather the tree based
method{,} {is} gaining all around popularity and soon {it} might
become the preferred method of choice.

\normalem
\begin{acknowledgements}
MS thanks P. Savicky for making available the decision plot of the
simulated MAGIC data. {This} helped in comparing the
decision plot of their simulated data from our program.
\end{acknowledgements}

\vs\vs

\appendix

\section{various machine learning methods}

\begin{figure}[!h]
\centering

\includegraphics[width=0.42\textwidth,angle=270,clip]{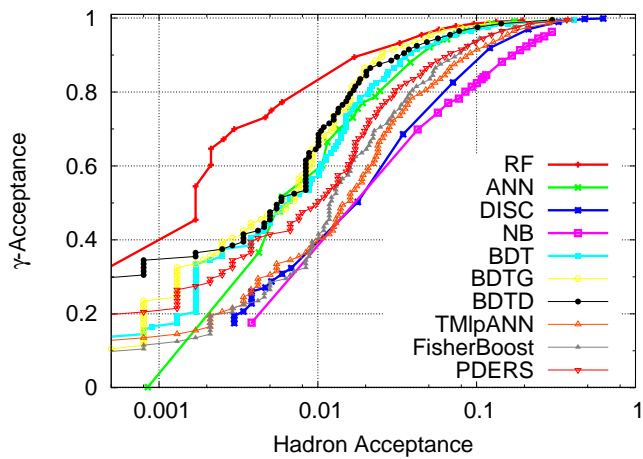}

\begin{minipage}{95mm}

\caption{\label{figa1} The decision-plot of various machine
learning methods.}\end{minipage}
\end{figure}

In addition to the {five} machine learning methods, various
machine learning methods from {the }TMVA package (\citealt{tmva})
were tested and their resultant decision-plot is presented.
Various machine learning methods are as follows: Boosting Decision
Tree{ (BDT}), BDT with gradient boost{ (BDTG}), BDT with
decorrelation{ (BDTD)} + Adaptive Boost, TMlpANN (ROOT's own ANN),
Fisher Boost (Linear discriminant with Boosting) {and }Probability
Density Estimator Range-Search{ (PDERS}). For all these methods,
the default settings given by the TMVA developers were used. It is
clear from the decision-plot (Fig.~\ref{figa1}) that {the }{RF}
method outperforms all the other methods.

\vs\vs

\small \setlength{\itemindent}{-3mm} \setlength{\itemsep}{-0.5mm}
\setlength{\baselineskip}{4.7mm}

\bibliographystyle{raa}
\bibliography{mridul-mlearning}

\end{document}